# Convolutional Neural Networks with Radio-Frequency Spintronic Nano-Devices

Nathan Leroux[1*], Arnaud De Riz[1], Dédalo Sanz-Hernández[1], Danijela Marković[1], Alice Mizrahi[1*] and Julie Grollier[1]

[1]Unité Mixte de Physique CNRS/Thales, Université Paris-Saclay, 91767 Palaiseau, France

[*]nathan.leroux@cnrs-thales.fr, alice.mizrahi@thalesgroup.com

Convolutional neural networks [1,2] are state-of-the-art and ubiquitous in modern signal processing and machine vision. Nowadays, hardware solutions based on emerging nanodevices are designed to reduce the power consumption of these networks. This is done either by using devices that implement convolutional filters and sequentially multiply consecutive subsets of the input, or by using different sets of devices to perform the different multiplications in parallel to avoid storing intermediate computational steps in memory. Spintronics devices are promising for information processing because of the various neural and synaptic functionalities they offer. However, due to their low OFF/ON ratio, performing all the multiplications required for convolutions in a single step with a crossbar array of spintronic memories would cause sneak-path currents. Here we present an architecture where synaptic communications are based on a resonance effect. These synaptic communications thus have a frequency selectivity that prevents crosstalk caused by sneak-path currents. We first demonstrate how a chain of spintronic resonators can function as synapses and make convolutions by sequentially rectifying radio-frequency signals encoding consecutive sets of inputs. We show that a parallel implementation is possible with multiple chains of spintronic resonators. We propose two different spatial arrangements for these chains. For each of them, we explain how to tune many artificial synapses simultaneously, exploiting the synaptic weight sharing specific to convolutions. We show how information can be transmitted between convolutional layers by using spintronic oscillators as artificial microwave neurons. Finally, we simulate a network of these radio-frequency resonators and spintronic oscillators to solve the MNIST handwritten digits dataset, and obtain results comparable to software convolutional neural networks. Since it can run convolutional neural networks fully in parallel in a single step with nano devices, the architecture proposed in this paper is promising for embedded applications requiring machine vision, such as autonomous driving.

## Introduction

Convolutional Neural Networks have widely contributed to the success of Artificial Intelligence since LeNet-5 [3] outperformed other Deep Neural Networks [2] for handwritten digits recognition. Due to their invariance to translations and local distortions, Convolutional Neural Networks not only excel in image recognition, but also in signal processing [1], and they are at the core of Artificial Intelligence applications like Generative Adversarial Networks [4]. For this reason, it is crucial that novel technologies, whose goal is to decrease the energy consumption of artificial intelligence, implement Convolution Neural Networks efficiently [5–7]. Hardware developed to accelerate the multiply-and-accumulate operations of neural networks, such as crossbar arrays of memristors, initially focused on realizing fully-connected layers wiring all the input neurons to all the output neurons with independent synaptic weights [8,9], an operation that they can achieve in a single step. Convolution layers, however, have a different operating principle. To perform a convolution on input data, multiply-and-accumulate operations with fixed synaptic weights (filters) are applied to consecutive subsets of the input. This process is often performed sequentially, as the filter is sled over neighboring inputs, as illustrated in Figure 1(a). In addition, the convolution operation has to process several input channels, and different filters need to be applied in order to compute different output features. The whole process has therefore a strong sequential character that requires storing intermediate computation steps in memory, which, for convolutions, has a prohibitive cost in terms of energy consumption, speed, and area. Finding ways to eliminate this sequential nature and implement convolutional neural networks in a fully-parallel manner, so that they can process their inputs in a single step, is therefore of great interest.

A wide span of research investigates hardware neuromorphic implementations of convolutional layers using CMOS [10–12], memristor crossbar arrays [13–17] as well as optics and photonics [18–20]. A particular effort aims at unfolding each convolutional layer into a sparse matrix of synaptic weights and mapping it to a crossbar array of memories to process convolutions fully in parallel [10,13,14,17]. Spintronics, whose memory, multifunctionality and dynamics are attractive for information processing is another technology actively studied for neuromorphic computing. Spintronic devices present advantages for the integration since they are compatible with CMOS and the same technology can be used to implement both artificial neurons and synapses [21,22]. Research shows that arrays of spintronic memories are promising for associative memories [23], spiking neural networks [24] and convolutional neural networks with time-domain computing [25]. However, implementing in parallel all the multiply-and-accumulate operations of a convolution requires extremely large crossbar arrays. Because of the small OFF/ON ratio of spintronic memories [26], such a crossbar array of spintronic memories for parallel convolutions would suffer from sneak path currents [27], so a different approach is required.

A way to mitigate this issue is to use spintronic devices that communicate through radio-frequencies such that the path of the information is not solely determined by electrical connectivity, but also constrained by selectivity in the frequency domain [28–30]. Frequency selectivity can be achieved by encoding the input information into radio-frequency signals of different frequencies, and using spintronic nano-resonators as artificial synapses that rectify frequency specific signals [31,32]. It has been demonstrated that chains of spintronic resonators can natively implement multiply-and-accumulate operations on radio-frequency (RF) signals [30,31]. This is promising to directly classify radio-frequency signals sensed from the environment with an antenna, as well as to establish communication between layers of neurons through RF signals rather than through wiring only. Furthermore, the RF multiply-and-accumulate operation has been demonstrated experimentally in a small system [32], and a simulated fully-connected perceptron has achieved classification as well as a software perceptron on an image benchmark of handwritten digits [31].

In the present paper, we show that chains of spintronic nano-resonators are suitable for the implementation of convolutional layers in deep neural networks based on radio-frequency communications between layers of neurons and synapses. First, we show how chains of spintronic resonators can implement convolutions on different sets of input RF signals presented sequentially as in Figure 1(a), and we show that this method integrates selectivity in the frequency domain. Then we show that it is possible to achieve these convolutions in a single step, with different chains implementing different multiply-and-accumulate operations, thus enabling ultrafast computation. We present how the resonators can be spatially arranged as a matrix of weights of an unfolded convolution, and propose a spatial arrangement that does not suffer from the sparsity specific to this type of matrices. We explain how this architecture can operate convolutions to extract different features in parallel, and how to assemble the spintronic chains performing MAC operations on incoming RF signals with spintronic nano-oscillators that emulate neurons and emit RF signals [11,33–40] in order to build deep neural networks. We also show that it is possible to train simultaneously all the spintronic resonators implementing the same filter coefficient by tuning them all at once. Finally, we simulate a full convolutional neural network made of these RF spintronic nano-devices and demonstrate an accuracy of 99.11 % on the Mixed National Institute of Standards and Technology (MNIST) handwritten digits dataset, the same accuracy obtained for a software network with an equivalent architecture.

# I. Radio-Frequency multiplications for spintronic convolutions

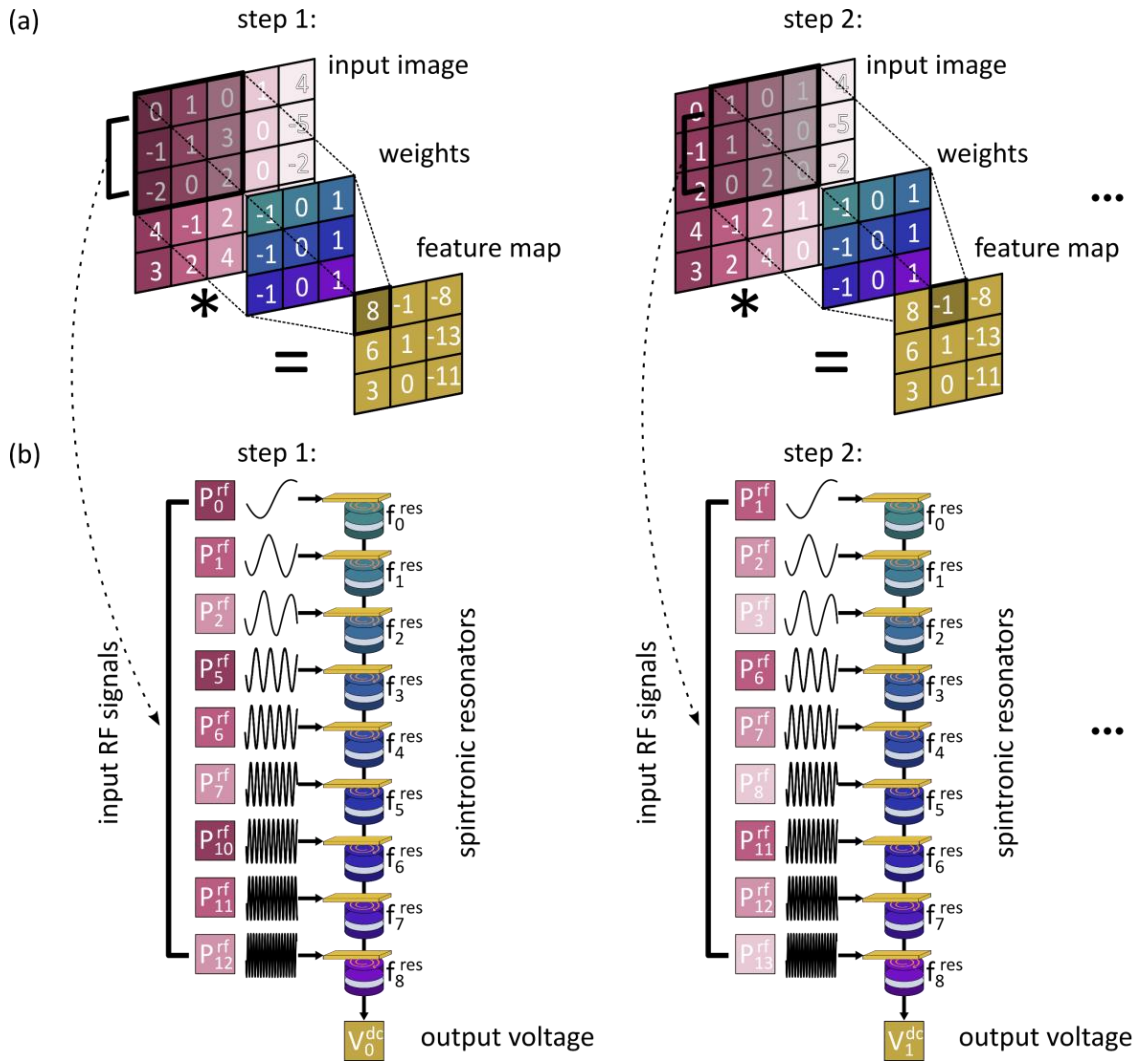

Figure 1: (a) Example of a 2D convolution with an input image of size 5X5 and a filter of size 3X3. Each element of the output feature map is the sum of the element-wise matrix multiplication (multiply-and-accumulate operation) between a subset of the input image and the weights of the kernel. The colors of the input image pixels indicate to which RF input signal they correspond in (b). (b) Schematic of the corresponding sequential convolution with RF signals and a chain of spintronic resonators. At each step the microwave powers of the input signals correspond to a mapping of a subset of the input image. The RF signals are injected to resonators through field-lines, represented by yellow stripes. Spintronic resonators are represented with colors corresponding to their matching weights, themselves represented in (a). At each step, the output voltage is a multiply-and-accumulate operation between the input microwave powers and the weights encoded into the resonance frequencies of the resonators.

In convolutional layers, the input image is convolved with multiple filters. The specificity of convolutions is their efficiency to extract spatial features, and each filter aims at extracting a different pattern present in the input image. To do so, each filter (which is often much smaller than the input image) slides over the input image, and at each position applies a multiply-and-accumulate operation to the corresponding image subset (see Figure 1(a)). Then the outputs, also called feature maps, store the result of the corresponding multiply-and-accumulate matrix operations (the sum of the elements of an element-wise matrix multiplication between the filter and a subset of the input image). In this section, we show how to perform these different multiply-and-accumulate operations sequentially using RF encoded inputs and a single chain of spintronic resonators for each filter. A parallelized architecture is presented in the next section.

Figure 1(b) shows a chain of spintronic resonators performing multiply-and-accumulate operations of a convolution. First, the intensity values of the input image pixels are mapped to RF powers corresponding to the pixel values of the image. Then, at each step, the corresponding subset of the input image is injected into the chain of resonators. For instance, during the first step an RF signal is injected into the first diode ($f_0^{res}$) with a power $P_0^{rf}$ and a frequency $f_0^{rf}$ corresponding to the first pixel. During the second step, the power in the same diode is changed to $P_1^{rf}$, corresponding to the second pixel. Each RF signal is injected through an individual field-line to one of the spintronic resonators of the chain [41]. Each resonator has a resonance frequency close to the frequency of its input RF signal. The spintronic resonators are employed in a "spin-diode" mode in which they filter and rectify the RF signals that they receive (see Methods for more details) [27,28]. Here, the rectification is caused by the mixing between the alternative current induced in the resonator by the current injected in the field-line (through capacitive or inductive effects), and the resistance oscillation of the resonator induced by magneto-resistive effects as the magnetization driven by the alternative field oscillates. Since the resonators of the chain are connected electrically, if we used the same RF frequency for all the inputs signals, each resonator could mix with signals induced by different field-lines and rectify them, hence causing crosstalk issues. On the contrary, here we choose a different frequency for each input signal to ensure that each resonator only mixes with the RF signal transmitted by its field-line, and thus rectifies the proper input signal. The dc voltages produced by the resonators are summed because they are electrically connected in series. The total voltage of such a chain of resonators is a multiply-and-accumulate operation between the input microwave powers and synaptic weights that are encoded in the resonance frequencies of the resonators [31,32]. This operation is described in detail in the Methods section. In this implementation, the difference between the input RF frequency and the resonance frequency of the resonators implements the weights of the convolutional filter. Since in a convolution, each multiply-and-accumulate operation requires the same set of weights, these resonance frequencies are left unchanged between the different steps. Then, at each step of the convolution, the voltage of the chain encodes a different element of the output feature map.

This method is straightforward, but it has the defect of being sequential. Doing these operations one after the other is costly in memory because it requires to store all the elements of the output feature map between two convolutional layers, and it slows down computing since it requires approximatively as many steps as there are pixels in its input images, versus a single one for parallel convolutions [16]. In the next section we describe how to operate RF convolutions fully in parallel.

## II. Fully-parallel architecture for Radio-Frequency convolutions

To perform all multiply-and-accumulate operations in parallel, we propose a novel architecture using multiple chains of spintronic resonators. This architecture is represented in Figure 2(b). Unlike the sequential method, here we simultaneously send all the elements of the input image to the resonators. Each pixel is mapped to a different RF signal with power proportional to its value, which is simultaneously injected into several spintronic resonators, each belonging to a different chain. A single field-line corresponds to a row in Figure 2(b). The different resonators of the same chain are arranged in a column and correspond to different coefficients of the filter. For instance, the resonators of the first column correspond to the coefficients $w_{1,1}$, $w_{1,2}$, $w_{2,1}$, and $w_{2,2}$ (in that order) of the filter, as it is represented in the left-hand side of Figure 2(a). The resonators of the second column correspond to the coefficients $w_{1,0}$, $w_{1,1}$, $w_{1,2}$, $w_{2,0}$, $w_{2,1}$, and $w_{2,2}$ of the filter, as it is represented in the right-hand side of Figure 2(a). The resonators of the fifth column correspond to all the coefficients of the filter, from the $1^{st}$ to the $9^{th}$. The resonators of the first column correspond to the coefficients $w_{1,1}$, $w_{1,2}$, $w_{2,1}$, and $w_{2,2}$ of the filter. The resonance frequencies of resonators in a row all match the corresponding input RF signal they need to rectify, but are not identical as they encode different synaptic weights: as the filter is sled, the same input gets multiplied by a different weight (see Figure 1(a) and Figure 2(a)). In Figure 2(b) each spintronic resonator is represented with a color that corresponds to one of the synaptic weights, which themselves are represented in Figure 2(a).

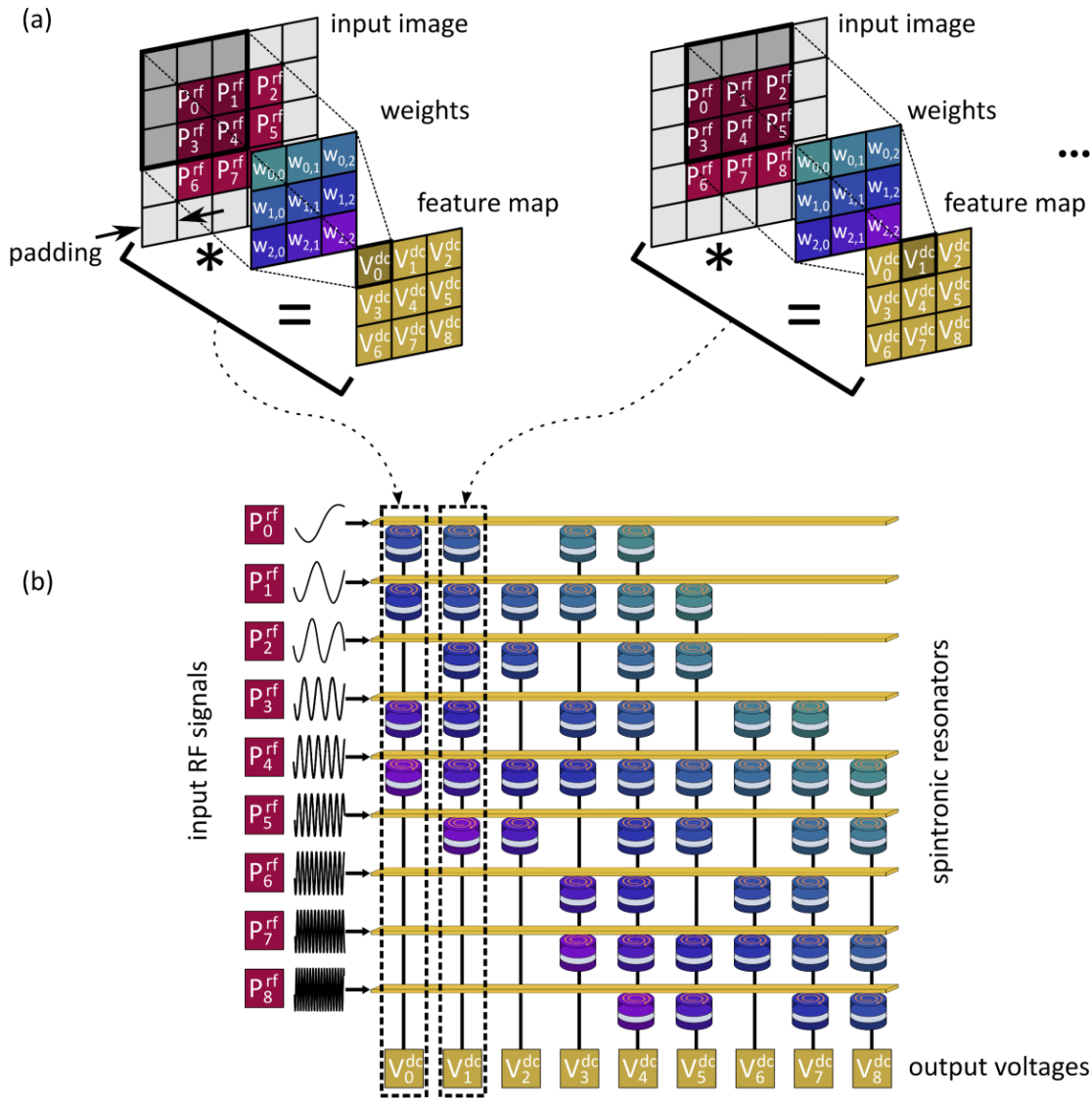

Figure 2: (a) Example of a 2D convolution with an image of size 3X3, a padding of 1 and a filter of size 3X3. Padding of 1 means that the image is padded at its outer edge with 1 layer of zeros. Each element of the output feature map is the sum of the element-wise matrix multiplication (multiply-and-accumulate operation) between a subset of the input image and the weights of the filter. (b) Schematic of the corresponding parallel convolution with RF signals and multiple chains of spintronic resonators. RF input signal powers are mapped to the image pixels. The RF signals are injected to resonators through field-lines, represented by yellow stripes. Spintronic resonators are represented with colors corresponding their matching weights, themselves represented in (a). Resonators encoding the same weights are aligned in diagonal. The voltage of each chain is a multiply-and-accumulate operation between a subset of the input microwave powers and synaptic weights encoded in the resonators' resonance frequencies.

We have to tune the resonance frequencies of the resonators to change the synaptic weights they implement in order to train the network. In Figure 3(a-b), all resonators implementing the same filter coefficient are represented with the same color. In the crossbar architecture studied until now, all the resonators that encode the same synaptic weights are aligned in a diagonal (see Figure 3(a)). Alternatively, as shown in Figure 3(b), we can change the spatial arrangement of the spintronic resonators, to implement a compacter architecture in which the resonators implementing the same filter coefficient are aligned in a column. Since they are aligned, we can tune simultaneously the resonators coding for the same synaptic weight with a single write-line. Write-lines provide an electrical control of synaptic weights either by changing the state of memristors placed above each spintronic resonator as it was done in [42] or in [43], or by switching the magnetization of spintronic resonators between two states [44–49] such as in binary neural networks [50–54]. Independently of the control method, a physical implementation of a network with the proposed architecture can be trained with a number of field-lines that does not scale with the number of devices, but only with the number of synaptic weights per filter.

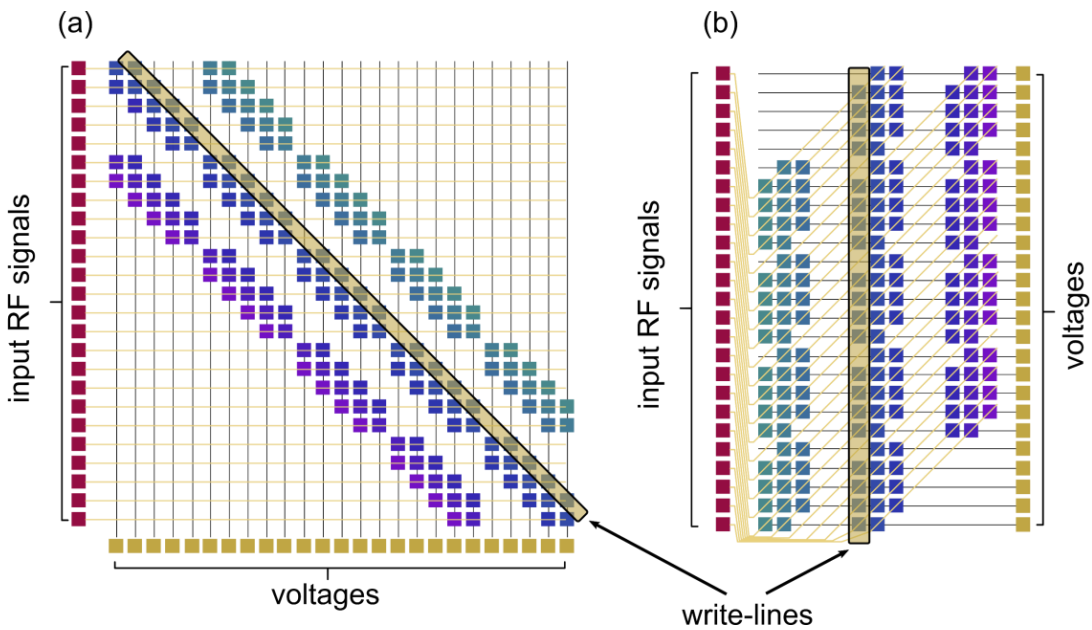

Figure 3: Schematics of a parallel convolution with an input image of size 5X5, a padding of 1 and a filter of size 3X3 with RF signals and chains of spintronic resonators implementing the synaptic weights. To simplify the figure, spintronic resonators are represented by colored squares in central panels and field-lines carrying RF signals are represented by yellow lines. Golden shade stripes represent write-lines that can simultaneously tune multiple spintronic resonators electrically. (a) Architecture similar to a traditional crossbar with input lines perpendicular to output lines. Spintronic resonators implementing the same synaptic weights are aligned in diagonals. (b) Compact architecture is a spatial rearrangement of synaptic components: unlike in traditional crossbars, output lines are parallel to input lines. Spintronic resonators implementing the same synaptic weights are aligned in columns.

III. **Multi-layer convolutional neural network implementation with chains of resonators and spintronic nano-oscillators**

In the previous sections we presented convolutions with single channel images and only one filter per layer. In typical convolutional layers, the input image is convolved with multiple filters, that produce different feature maps. Each feature map becomes a different channel of the input image in the next layer, as shown in Figure 4(b). Filters are 3D tensors whose depth is equal to the number of channels of their input image. The pixel values in each of the $N_m$ different feature maps resulting from the convolution of an input image with $N_c$ channels with a filter of size $k \times k \times N_c$ are given by the formula:

$$z_{h,w,m} = \sum_{c=0}^{N_c-1} \sum_{j=0}^{k-1} \sum_{i=0}^{k-1} W_{i,j,c,m} x_{i+h,j+w,c} + b_m, \qquad (1)$$

where $W$ are the filter coefficients, $x$ the input pixel values, $b$ are the biases, $m$ is the feature map index, $h$ and $w$ are the height and width positions of the pixel in the feature map, $i$ and $j$ are the vertical and horizontal coordinates of the pixel in the filter, and $c$ is the input channel index. To make this operation fully parallel, additional resonators are employed as illustrated in Figure 4(a). To implement different channels, additional sets of RF signals (in blue) are employed, which are sent to additional sets of spintronic resonators connected in series with the resonators rectifying the RF signals from the first channel (in red). Additional filters are implemented by adding new chains of resonators after the chains of resonators implementing the first filter, as in the right-hand side of Figure 4(a). Similarly, multiple channels can be convolved with multiple filters with the compact architecture described in Figure 3(b).

Building deep neural networks requires transferring information between different layers of neurons and synapses. We propose using spintronic oscillators as artificial neurons emitting RF signals as inputs to each convolutional layer. It was already demonstrated that these oscillators can be used as artificial neurons thanks to their nonlinear dynamics [33,35,36]. Here, we leverage the nonlinear transformation between the input direct current to the oscillators and the output microwave power they produce [33,55,56] (see Methods). Using the fully parallel architecture, it is then possible to cascade the information between different convolutional layers: provided a voltage-to-current amplification, every direct voltage output of the convolution can be used to supply an artificial neuron of the next layer. Then the RF signals emitted by spintronic oscillators become the inputs of chains of spintronic resonators, these resonators rectify RF signals into direct voltage which supplies the next layer of spintronic oscillators, and so on. This architecture alternating between RF and DC signals is illustrated in Figure 4(b). In practice, CMOS components are needed to convert voltages to currents and amplify them to match the threshold current of spintronic oscillators.

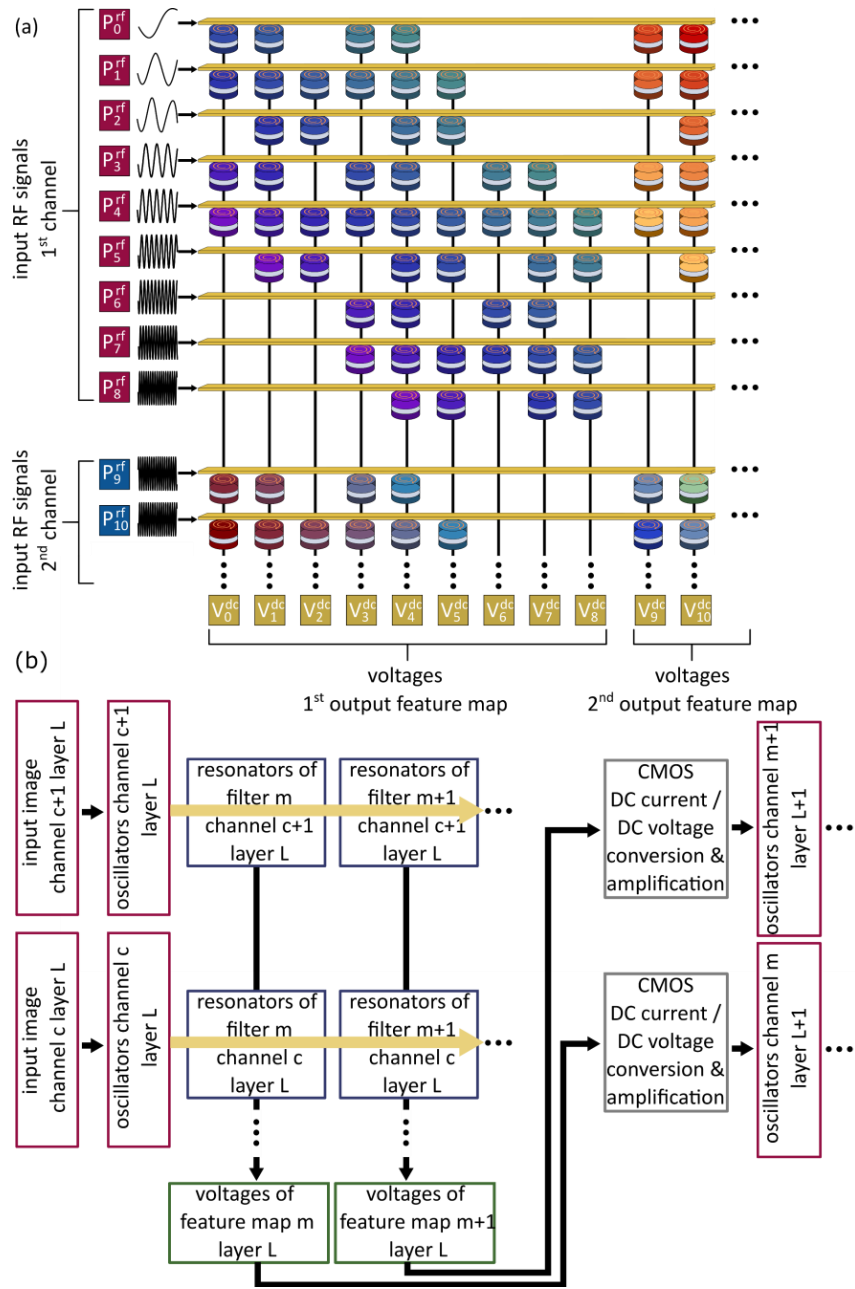

Figure 4: (a) Schematic of a parallel convolution with RF signals and chains of spintronic resonators implementing the synaptic weights similar to Figure 2(b) but with different channels and features. First (second) input channel pixels are mapped to a first (second) set of RF signals represented in red (in blue). Chains of spintronic resonators corresponding to the outputs of the second feature map are parallel to the first set of chains corresponding to the ouputs of the first feature map. (b) Diagram of a deep convolutional network with multiple channels and features with spintronic oscillators to emulate neurons and resonators to emulate synapses. Direct signal electrical connections are represented by black arrows while field-lines carrying RF signals are represented by yellow arrows.

As in the two previous sections, a different frequency is assigned to each oscillator. Here again, because we use field-lines to transmit RF signals locally to specific resonators and because we rely on a resonant effect, each artificial synapse only multiplies its corresponding input. Thus, our implementation can perform convolutions in parallel using multiple resonator chains with limited crosstalk that are problematic in most hardware neural networks implementations, like sneak-path in crossbar memristor arrays [8,27,57–59].

## IV. Handwritten digits classifications with a Convolutional Neural Network implemented on spintronic nano-oscillators and resonators

In this section, we simulate a network with spintronic oscillators as neurons and spintronic resonators as synapses based on the proposed convolutional architecture. The goal is to prove that chains of resonators can calculate convolutional operations with high accuracy, that it is possible to tune the convolutional filter coefficients by tuning the resonance frequencies, and to demonstrate the capacity of spintronic oscillators to implement activation functions in such networks.

We benchmark our network on the standard MNIST dataset. It consists of 28 X 28 pixel images of handwritten digits. The topology of our network is shown in Figure 5(a); 32 filters of 5 X 5 with stride 1 and padding 1 for the first convolutional layer, a max-pooling of size 2 X 2 and stride 2, a layer of spintronic oscillators as activation functions, 64 filters of 5 X 5 with stride 1 and padding 1 for the second convolutional layer, a second max pooling of size 2 X 2 and stride 2, a second layer of spintronic oscillators as activation functions, a fully connected layer of size 1600 X 10, and a softmax layer in the end. The physical analytical models used to simulate spintronic devices are detailed in the Methods. It has already been proven experimentally in [27, 28] that fully-connected layers can be implemented by chains of spintronic resonators, applying multiply-and-accumulate operations on RF encoded inputs and the same operating principle is assumed here. Max-pooling layers and the softmax layer are assumed to be implemented by more classical technologies such as CMOS circuits.

To train the network, we use 60,000 images for training and 10,000 for testing. At each training iteration, a batch of 20 images is presented to the network, the output of the network is computed with the softmax layer, and the cost function is the Cross-Entropy Loss [60]. We use the backpropagation algorithm [2] and Adam optimizer [61] of the software PyTorch to train the network.

In our implementation, the synaptic weights depend both on the frequency of the input and the resonance frequency of the resonators (see Eq. 5 in Methods). The input frequency is kept fixed and the trained parameter is the resonance frequency of the resonators. An additional constraint arises in the case of parallel convolutions due to weight sharing. Indeed, resonators corresponding to the same filter coefficient have to implement the same synaptic weight even if they receive input signals different frequencies. In order to ensure that this is the case, our training algorithm updates the resonance frequency of each resonator with a function that depends both on the frequency it receives, and a trainable parameter $\zeta_{i,j,c,m}$ learned through backpropagation that corresponds to its filter coefficient:

$$f^{res}_{i,j,c,h,w,m} \leftarrow f^{RF}_{h+i,w+j,c}(1 - \zeta_{i,j,c,m}). \tag{2}$$

This expression indicates that when multiple spintronic resonators are tuned simultaneously with a single write-line in a hardware implementation, the resonance frequency update of each resonator should scale with its input signal frequency. In the Methods section, we demonstrate that using this expression, chains of resonators voltages correspond to convolution outputs, described by Eq. (1). For the fully-connected layer, we simply set the resonance frequencies as trainable parameters learned through backpropagation, as it was done in [31]. The learning rate is $10^{-4}$.

In experimental hardware implementations, signals should be amplified between a layer of spintronic oscillators and a layer of spintronic resonators. Therefore, in our simulations we introduce an amplification factor for each synaptic layer. Amplification factors are set as trainable parameters and are trained through backpropagation. Adjusting these parameters during training balances the fact that spintronic resonators can only be tuned within a finite range of synaptic weights.

We plot in Figure 5(b) the learning curve of the network. We see that at the end of the training, our spintronic network (red solid line) classifies handwritten digits with 99.11 % accuracy, as good as the 99.16 % accuracy we found with a software convolutional neural network with the same architecture (solid blue line). The difference in accuracy is smaller than the standard deviation, represented in shaded area in Figure 5(b). In addition, the classification results are higher than with a Multi-Layer-Perceptron [3], which indicates that the advantages of convolution are preserved with our RF spintronic network. These results show the feasibility of convolutions with RF signals and spintronic resonators.

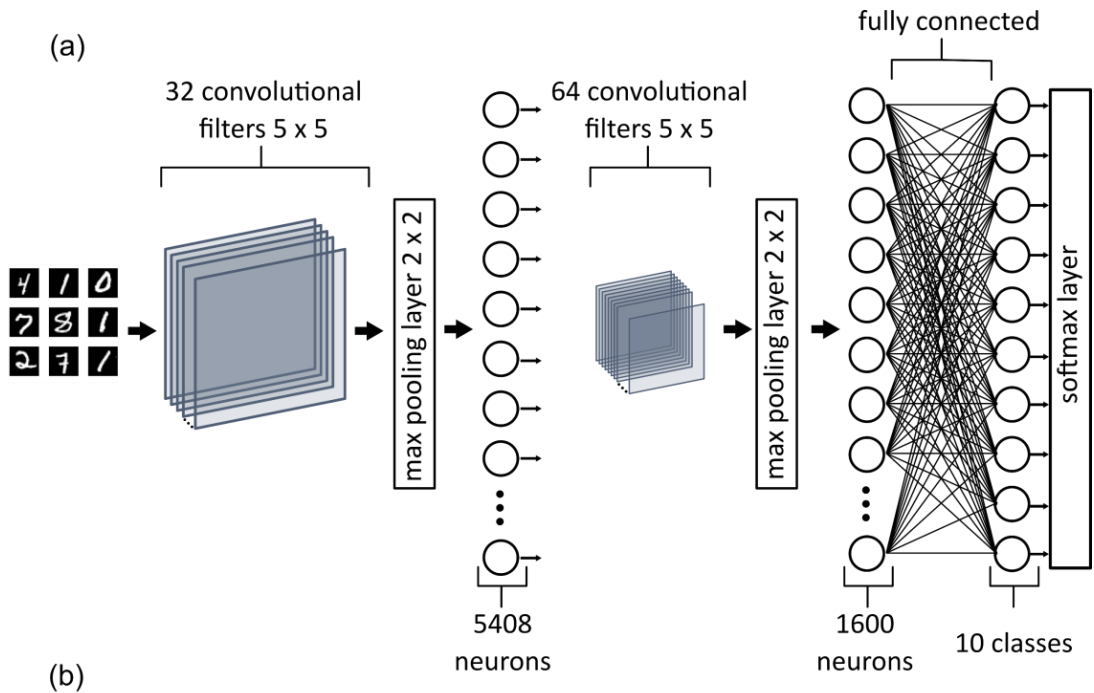

Figure 5: (a) Neural network architecture used to classify images of the MNIST dataset. (b) Accuracy of image recognition measured in percentage of successful classification, versus the number of epochs (the number of time we present the training dataset to the network to train it). The red (blue) solid line shows the mean accuracy for a simulated RF spintronic devices network (a software neural network) for ten repetitions on the test dataset. In green (black) solid line the mean accuracy for a simulated RF spintronic devices network (a software neural network) on ten repetitions on the training dataset. Shaded areas represent standard deviations.

## Conclusion

Using chains of spintronic resonators as artificial synapses rectifying RF signals into dc voltages, and nano-oscillators as nonlinear activation functions converting direct currents to RF signals, our system cascades information between different neural layers and performs the different multiplications of convolutions fully in parallel. In this paper, we showed how we can use chains of resonators to rectify multiple RF signals to make convolutions. We described the concept through a sequential convolution with a single chain of resonators, and then we showed how it can be extended to operate convolutions in parallel with multiple chains. This parallelization provides a tremendous processing time reduction: a sequential convolution requires approximatively as many steps as there are pixels in its input images, versus a single one for parallel convolutions. We have highlighted that with the two different proposed architectures, a single write-line can tune simultaneously many resonators to adjust their synaptic weights. The number of resonators adjustable by the same write-line is equal to the number of positions the filters take in the convolution. This can reduce the complexity and time required to train such hardware neuromorphic architecture. Since we use a resonance effect and field-lines that locally inject RF signals to spintronic resonators, each synapse only multiplies its corresponding neural input. This could suppress crosstalk problems that are common in neuromorphic architectures, like sneak-path in crossbar memristor arrays, and help to use a large number of devices in parallel to perform convolutions in a single step. Using spintronic oscillators as RF emitters emulating neurons, we show that chains of resonators can rectify neural signals, and that these rectification voltages can supply another layer of spintronic oscillators, hence enabling direct connectivity between different neural layers. We have demonstrated the performance of such network on the MNIST dataset though physical simulations of these RF spintronic devices and obtained 99.11 % accuracy. Spintronic oscillators and resonators are similar devices that require the same materials and are both compatible with CMOS technology. Moreover, as these components can be downscaled to 20 nm, our architecture can contain a very large number of devices on a small surface. The density of devices is critical in a neural network, both to reduce power dissipation and to allow parallelism. The computing parallelism also limits the inference latency, which is crucial for modern embedded applications such as autonomous driving. In conclusion, this work opens new avenues to very large spintronic neuromorphic networks able to efficiently perform convolutions.

## Acknowledgments

This project has received funding from the European Research Council ERC under Grant bioSPINspired 682955, from the European Union's Horizon 2020 research and innovation program under grant agreement RadioSpin No 101017098 (project) and the French Ministry of Defense (DGA). The authors thank Erwann Martin and Jérémie Laydevant for their scientific support.

## Methods

For the numerical simulations of handwritten digit classification, the physical response of hardware-based convolutional layers, nonlinear layers and fully connected layers is simulated based on well-established physical models of spintronic devices as detailed below.

### Spintronic oscillators simulations

The output of each spintronic oscillator is computed with the universal auto-oscillator theory [55] in the case of a spin-torque nano-oscillator The normalized magnetization power is

$$p = \begin{cases} \dfrac{I_{dc}/I_{th} - 1}{I_{dc}/I_{th} + Q} & if\ I_{dc} > I_{th} \\ 0\ if\ I_{dc} \leq I_{th} \end{cases} \tag{3}$$

with $I_{dc}$ the direct current, and Q is the nonlinear damping coefficient. In this paper we chose $Q = 2$ according to experimental works [28,34,62]. In the handwritten digits demonstration we chose a threshold current of 2 mA because it is coherent with experimental works [33,42] and we clamp each direct current input to 8 mA because in practice, these nano-devices can be damaged above a particular current [56]. Since the threshold current density is typically around $10^{11}$ A/m², in the future the threshold current of these oscillators can be decreased to tens of μA with lateral size reduction to tens of nanometers. Moreover, projections show that size reduction could also reduce threshold current density down to $10^{10}$ A/m² [63].

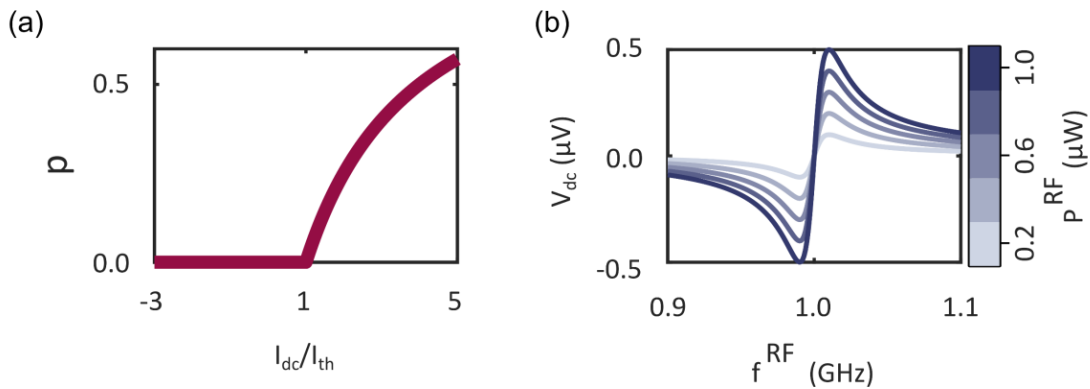

Figure 6: a) Normalized magnetization oscillation power of a Spin-Torque Nano-Oscillator versus the ratio between direct current and threshold current. Plot realized using Eq. 3 with $Q = 2$. b) Voltage rectified by spin-diode effect versus microwave frequency for different microwave powers $P^{RF}$ = 0.2, 0.4, 0.6, 0.8 and 1 μW. Plot realized using Eq. 4 with $f^{res} = 1$ GHz, $\alpha = 0.01$, and $\mathcal{A} = 1$ μV/μW.

Most spintronic oscillators have a frequency dependence with input direct current [55,64,65]. We want to avoid this effect because, as we see in Eq. 5, the weights also depend on microwave frequencies, and we cannot allow the weights to depend on the input values. To avoid these frequency shifts, oscillators with compensated magnetic anisotropy can be used [62,66].

**Spintronic resonators simulations**

Due to spin-diode effect, the voltage rectified by each spintronic resonator is:

$$V(P^{RF}, f^{RF}, f^{res}) = P^{RF} f^{RF} \frac{f^{RF} - f^{res}}{\alpha^2 f^{res^2} + (f^{RF} - f^{res})^2} \mathcal{A}. \qquad (4)$$

This expression comes from the experimentally validated [67,68] auto-oscillator model [55]. Here $P^{RF}$ is the emitted power of an oscillator, $f^{RF}$ its frequency, $f^{res}$ is the resonance frequency of a spintronic resonator, $\alpha$ the magnetic damping of the material used for the spintronic resonators, and $\mathcal{A}$ a scaling factor. Writing Eq. 4 we neglect the frequency-symmetric Lorentzian component of the voltage and keep only the anti-symmetric Lorentzian component [67]. This approximation is useful because it lets the synaptic weight be either positive or negative, and is valid because in practice it is possible to tune the ratio between the symmetric and the anti-symmetric components by changing the orientation of a magnetic field [69,70].

Then each rectification of a RF signal with a spintronic resonator is a multiplication operation on the microwave power $P^{RF}$, with a weight

$$W(f^{RF}, f^{res}) = f^{RF} \frac{f^{RF} - f^{res}}{\alpha^2 f^{res^2} + (f^{RF} - f^{res})^2} \mathcal{A}. \qquad (5)$$

We see that because of the anti-Lorentzian shape of Eq. 5, we can tune the amplitude and the sign of the synaptic weight by tuning the resonance frequency $f^{res}$. In this paper we chose a magnetic damping coefficient $\alpha = 0.01$, which is the magnetic damping of permalloy [71], and an arbitrarily chosen amplification factor $\mathcal{A} = 1$ µV/µW. For the neural network, the resonance frequencies are randomly initialized between 1 and 2 GHz for each layer.

**Convolution specific multiply-and-accumulate operations with chains of spintronic resonators**

In this subsection we demonstrate that with the correct choice of resonance frequencies, the voltages of our chains of spintronic resonators are convolution outputs like Eq. 1. From Eq. 4, we know that in a convolutional framework the voltage of each resonator inside each chain is

$$V_{i,j,c,h,w,m} = P^{RF}_{h+i,w+j,c} f^{RF}_{h+i,w+j,c} \frac{f^{RF}_{h+i,w+j,c} - f^{res}_{i,j,c,h,w,m}}{\alpha^2 {f^{res}_{i,j,c,h,w,m}}^2 + \left(f^{RF}_{h+i,w+j,c} - f^{res}_{i,j,c,h,w,m}\right)^2} \mathcal{A}. \qquad (6)$$

Replacing $f^{res}_{i,j,c,h,w,m}$ in this equation by the expression of Eq. 2, we find

$$V_{i,j,c,h,w,m} = P^{RF}_{h+i,w+j,c} \frac{\zeta_{i,j,c,m}}{\alpha^2\left(1-\zeta_{i,j,c,m}\right)^2 + \left(\zeta_{i,j,c,m}\right)^2} \mathcal{A}. \tag{7}$$

Since the total voltages of each chain of spintronic resonators is the sum of the voltages of each resonator of the chain, they are equal to

$$U_{h,w,m} = \sum_{c=0}^{N_C-1} \sum_{j=0}^{k-1} \sum_{i=0}^{k-1} P^{RF}_{h+i,w+j,c} \frac{\zeta_{i,j,c,m}}{\alpha^2\left(1-\zeta_{i,j,c,m}\right)^2 + \left(\zeta_{i,j,c,m}\right)^2} \mathcal{A}, \tag{8}$$

and we can identify this equation to Eq. 1, with $U_{h,w,m}$ equivalent to the outputs $z_{h,w,m}$, $P^{RF}_{h+i,w+j,c}$ equivalent to the inputs $x_{i+h,j+w,c}$, and $\frac{\zeta_{i,j,c,m}}{\alpha^2\left(1-\zeta_{i,j,c,m}\right)^2 + \left(\zeta_{i,j,c,m}\right)^2} \mathcal{A}$ equivalent to the filter coefficients $W_{i,j,c,m}$.